\documentclass[aps,prb,amsmath,amssymb,amsfonts,floatfix,nofootinbib,superscriptaddress,longbibliography,12pt,reprint,noeprint]{revtex4-2}

\usepackage{geometry}
\usepackage[english]{babel}
\usepackage{blindtext}
\usepackage{amsmath,amsfonts,amssymb}

\usepackage{graphicx}
\usepackage{float}
\usepackage{times}
\usepackage{color}

\usepackage{subfigure}
\usepackage{indentfirst}
\usepackage{setspace}
\usepackage{bm}% bold math

\usepackage[colorlinks,linkcolor=blue,anchorcolor=blue,citecolor=green]{hyperref}
\usepackage{CJKutf8}
\newcommand{\Tr}{\mathrm{Tr}}

\graphicspath{{pdf}}

\allowdisplaybreaks
\newcounter{amgg}
\newcounter{fan}
\newcounter{yc}

\begin{document}

\title{Entanglement dynamics of monitored noninteracting fermions on graphics processing units}
\author{Bo Fan (\begin{CJK*}{UTF8}{gbsn}范波\end{CJK*})$^*$}
\email{bo.fan@sjtu.edu.cn}
\affiliation{Shanghai Center for Complex Physics, School of Physics and Astronomy,
\\Shanghai Jiao Tong University, Shanghai 200240, China}
\author{Can Yin (\begin{CJK*}{UTF8}{gbsn}殷灿\end{CJK*})$^*$}
\email{yin\_can@sjtu.edu.cn}
\affiliation{Shanghai Center for Complex Physics, School of Physics and Astronomy,
\\Shanghai Jiao Tong University, Shanghai 200240, China}
\author{Antonio M. Garc\'ia-Garc\'ia}
\email{amgg@sjtu.edu.cn}
\affiliation{Shanghai Center for Complex Physics, School of Physics and Astronomy,
\\Shanghai Jiao Tong University, Shanghai 200240, China}
\vspace{0.2cm}
\date{\today}
\vspace{0.3cm}

\begin{abstract}
The description of the entanglement dynamics of monitored noninteracting fermions, including the existence of measurement-induced phase transitions (MIPTs), is a challenging problem with conflicting results in the literature. The mapping of the problem onto a non-linear sigma model (NLSM) indicates that relatively large lattice sizes are required to determine the nature of the entanglement entropy (EE) in the thermodynamics limit.  
Here we address this problem numerically for monitored noninteracting fermions with $U(1)$ symmetry. The use of graphics processing unit (GPU) techniques, even with outdated hardware, makes it possible to reach much larger lattice sizes ($L = 16384$ and $160\times160$ in one (1d) and two (2d) dimensions respectively) than in previous studies which enables us to characterize quantitatively the entanglement dynamics. In 1d, we show that in order to confirm the absence of a MIPT, for both projective and homodyne measurements, predicted by the NLSM it is necessary to reach $L \sim 10000$. In 2d, also as predicted by the NLSM, we observe for both protocols a MIPT at finite monitoring rate characterized by a scale invariant mutual information. The critical monitoring strength depends on the protocol while the critical exponent $\nu \approx 1.3$ governing the approach to the MIPT is similar in both cases. These features are not correctly predicted by the NLSM. Our results paves the way for a fully quantitative description of the entanglement dynamics of monitoring quantum systems.      

\end{abstract}

\maketitle
\def\thefootnote{*}\footnotetext{These authors contributed equally to this work}
\section{Introduction}
Measurement induced phase transition (MIPT) \cite{Li2018a,Skinner2019a,nahum2021a,ippoliti2021a,Li2019a,carisch2023,Szyniszewski2019a,Szyniszewski2020,Turkeshi2020a,Turkeshi2021,Turkeshi2022a,legal2023,Soares2024,Sieberer2025} in the entanglement entropy (EE) has attracted a lot of attention because of its relevance in both quantum information applications and the foundations of quantum mechanics. The existence of MIPT has been confirmed experimentally \cite{Noel2022a,Koh2022} though a more systematic description of its properties is still missing.
For many-body quantum systems, even in noninteracting theories, it is indeed a challenging problem to determine the conditions for the existence of MIPT either numerically or analytically. For free Dirac fermions in 1d with no disorder and nearest neighbors hopping, it has been shown analytically \cite{poboiko2023,fava2024}, by mapping the problem onto the NLSM \cite{wegner1979,efetov1980}, that projective measurements of the occupation number always lead to an EE that does not scale with system size, namely, there is no MIPT. 
However, also using the NLSM, a MIPT occurs \cite{fava2023} at a finite monitoring rate for Majorana fermions subjected to homodyne measurements of the parity  \cite{wiseman1993,collett1987,wiseman2014,fuwa2015} characterized by a quantum state evolution governed by a stochastic Schrodinger equation \cite{cao2019a,alberton2021a,carisch2023,ladewig2022}. The difference between the two results is traced back to the different symmetries of the NLSM. 
Intriguingly, the MIPT has striking similarities \cite{poboiko2023} with the single-particle Anderson metal-insulator transition in 2d \cite{anderson58,abrahams1979} belonging to the BDI \cite{fava2024} universality class.

Some numerical results for 1d Dirac fermions subjected to projective \cite{poboiko2023} and homodyne \cite{cao2019a} measurements implemented  by the quantum trajectory method \cite{molmer93,dalibard1992,dum1992,daley2014} have confirmed the prediction of the NLSM about the absence of a MIPT. By contrast, Refs.~\cite{buchhold2021a,alberton2021a} reported a MIPT employing a quantum jump \cite{warren1986,zoller1987,gleyzes2007,minev2019,plenio1998} protocol. The numerical results were supported by semi-analytical arguments based on a Keldysh field theory approach and noticing the similarities of this problem with the physics of vortex unbinding. All these results are obtained for relatively small $L < 1000$ lattice sizes and in Ref.~\cite{cao2019a} the EE partition does not scale with system size.
Currently, the growing consensus is that there is no MIPT in 1d for Dirac fermions, at least in the studied universality classes. 

A MIPT has been reported in 2d noninteracting non-topological  \cite{poboiko2023a,chahine2024} fermions, topological \cite{xiao2025} fermions and in fermionic quantum circuits \cite{Jian2023,Jian2022} using different measurement protocols. Adding weak interactions \cite{fazio2024,poboiko2024,Li2024,guo2025,muller2025,lumia2024} or extending the range of hopping \cite{minato2022,mueller2022,Fuji2020a}, can also induce a MIPT. The interplay of disorder and monitoring has been investigated in \cite{lunt2022,paul2024} and the role of multifractality at the MIPT in Refs.~\cite{Jian2023,poboiko2025}.

In this paper, we study numerically the entanglement dynamics of monitored noninteracting Dirac fermions, namely, fermions with $U(1)$ symmetry. By using graphics processing units (GPUs), we are able to reach much larger sizes ($L \geq 16000$ in 1d and $\sim 160\times160$ in 2d)  than in previous studies ($L \lesssim 2000$ in 1d and $\sim 60\times 60$ in 2d) in the literature \cite{poboiko2023,chahine2024,cao2019a,alberton2021a}. Based on the mentioned analogy with the 2d  \cite{mackinnon1983,eilmes1998} and 3d Anderson model \cite{schreiber1991,garcia2007dimensional,rodriguez2011multifractal}, we believe the studied lattice sizes are enough for a fully quantitative description of the dynamics. \\
\section{The Model and Methods} 
The Hamiltonian of the system in 1d is given by, $
H = J \sum_{i=1}^{L} [ c_{i}^\dagger c_{i+1} + c_{i+1}^\dagger c_{i} ]$,
where $c_i$ and $c_i^\dagger$, $i=1,2,\cdots L$ are the annihilation and creation operators for fermions at site $i$ and we set $J=1$. We impose periodic boundary conditions $c_{i}=c_{i+L}, c_{i}^\dag=c_{i+L}^\dag$, the filling rate is $N/L=1/2$, with $N$ the number of fermions, and the initial state is $|\psi\rangle_0=|010101\cdots\rangle$ with $0,1$ site occupation numbers. We consider projective (PM) and homodyne measurements protocols, the latter termed quantum state diffusion (QSD), see Appendix \ref{app:protocol} for further details. For both protocols, the observable being measured is the occupation number $n_i =c_i^\dag c_i, i=1,2, \cdots L$. 
 In the PM protocol, projective measurements of the occupation number of individual sites, chosen randomly, occur at random times whose frequency is governed by a Poisson distribution depending on a parameter $\gamma$. The outcome of this measurement results in the projection of the wavefunction to one eigenstate of $n_i$ with eigenvalue $0, 1$. The probability of each of these two outcomes is given by Born's rule which in our case is just $1/2$. By contrast, in the QSD protocol, all sites are weakly measured at each time step. The state evolution is governed by a stochastic equation with a Gaussian noise of zero average and variance $\gamma dt$ where $dt$ is the time step.
 
Since the full system is quadratic even in the presence of monitoring,
the application of the Wick's theorem enables us to express any observable in terms of the $L\times L$ correlation matrix $D$ given by $D_{ij}(t)\equiv \langle \psi(t)|c_i^\dag  c_j|\psi(t)\rangle$, $i,j=1,2,\cdots, L$ where the explicit expression of the time dependence $|\psi(t)\rangle$ for each protocol is in Appendix \ref{app:protocol}. 
The EE is defined as $S_A(t)=\Tr_A(\rho(t) \ln(\rho(t)))$. For the moment, we will use as
subsystem $A=\{1,2,\cdots L/2\}$ ($\rho(t)=|\psi(t)\rangle \langle \psi(t)|$ is the density matrix) is $
	S_A(t) = -\sum_{i=1}^{L/2} \left( \lambda_i(t) \ln(\lambda_i(t)) + (1 - \lambda_i(t)) \ln(1 - \lambda_i(t)) \right)$
where $\lambda_i$ are the eigenvalues of $D_{ij}$.
The scaling with system size of the EE after the saturation time is an indicator of a MIPT. However, following Ref.~\cite{poboiko2023}, it is more convenient to expand the EE as a series of cumulants of the occupation number \cite{levitov2009,poboiko2023}.
 Perturbatively, it was found in Ref.~\cite{poboiko2023,levitov2009} that it is only necessary to consider the second cumulant proportional to the density-density correlation function $C(x-y,t)$,
  \begin{equation}
     S_A(t)\approx \frac{\pi^2}{3}\int_{0}^{L/2} dx\int_{0}^{L/2} dy \,C(x-y,t).
  \end{equation}
  For computational purposes, we use $C(r,t)=\overline{C(|x-y|=r,t)}$ where the overline denotes an average over all pairs of $(x,y)$ satisfying $r=|x-y|$, and also over quantum trajectories, namely, different realizations of the $|\psi(t)\rangle$ resulting from the monitored dynamics. In a lattice,  $C_{ij}(t)$, with $i,j$ denoting site indices, is obtained from the correlation matrix $D_{ij}(t)$ using Wick’s theorem,
 \begin{equation}
   C_{ij}(t)=-D_{ij}(t)D_{ji}(t)\qquad C(r,t)=\overline{C_{ij}(t)}, |i-j|=r
   \label{eq:Cr}
 \end{equation}
 where we only consider the case $i\neq j$. We will focus on long times $t \ge L/2$ where the EE has reached its saturation value. Later, we shall show explicitly that up to an overall prefactor, Eq.~(\ref{eq:Cr}) is still accurate \cite{poboiko2023,poboiko2023a} beyond the perturbative regime $\gamma \ll 1$.
  \section{Entanglement dynamics in 1d}
  We now proceed with the calculation of the EE for the 1d Hamiltonian above subjected to the PM and QSD protocols. 
  \subsection{Projective Measurement Protocol}
   For the PM protocol (see Appendix \ref{app:protocol} for further details), we recall that the NLSM prediction \cite{poboiko2023} for the BDI class \cite{fava2024} is that, $C(r,t\to\infty) \sim \exp(-r/l_{\rm cor})$ for $r \gg l_{\rm cor}$ and $C(r,t\to \infty) \sim 1/r^2$ for $r \ll l_{\rm cor}$, where $l_{\rm cor}$ is the correlation length. The asymptotic exponential decay implies that EE is in the area-law phase for any $\gamma > 0$. 
 Importantly, the correlation length \cite{poboiko2024,fava2024} 
  \begin{equation}
  	l_{\rm cor}\sim \frac{1}{\gamma}\exp\left(\frac{\sqrt{2}\pi}{2\gamma} \right)
  	\label{eq:RG}
  \end{equation} 
  increases exponentially with $1/\gamma$. Therefore, for $\gamma \ll 1$, it is challenging to confirm numerically the area-law phase because eventually $l_{\rm cor} \gg L$ so only a power-law decay, related to a {\it volume}-law phase, is observed. 
  However, the latter is a finite size artifact. 
A similar problem occurs in the 2d Anderson model \cite{mackinnon1983,eilmes1998} where it is necessary to reach $L^2 \sim 100\times 100$ to show the analogue result that, for any disorder, all eigenstates are exponentially localized.   
  
In order to proceed, we compute $C(r,t)$ in Eq.~\eqref{eq:Cr} for $t \ge L/2$, so that EE has reached its saturation value, by using GPU techniques. To be more explicit, in the QSD protocol, we implement the bulk of our matrix operations on the GPU (device) using C++ and CUDA platforms, reserving the CPU (host) for only lightweight tasks that do not justify additional kernel launches. The PM protocol relies on MATLAB's GPUArray framework to treat the GPU as an accelerator of the CPU workload, which is similar to CUDA when no fine-grained control over memory allocation or thread organization is needed.

We found that evolving the EE on the GPU is significantly faster and more efficient than on the CPU. A single NVIDIA A100 GPU outperforms roughly one hundred CPU cores (e.g., Intel Xeon 8362, $2.8\,\mathrm{GHz}$ base / $3.6\,\mathrm{GHz}$ boost), so a workstation with eight A100 cards can match the performance of about one thousand CPU cores. We checked that CUDA and MATLAB code provided similar performance for basic matrix manipulations. The calculation, for the PM protocol, up to $L = 8192$, was carried out with only two A100 cards running for about four months.

\begin{figure}[!htbp]
	\centering
	\subfigure{\includegraphics[width=7cm]{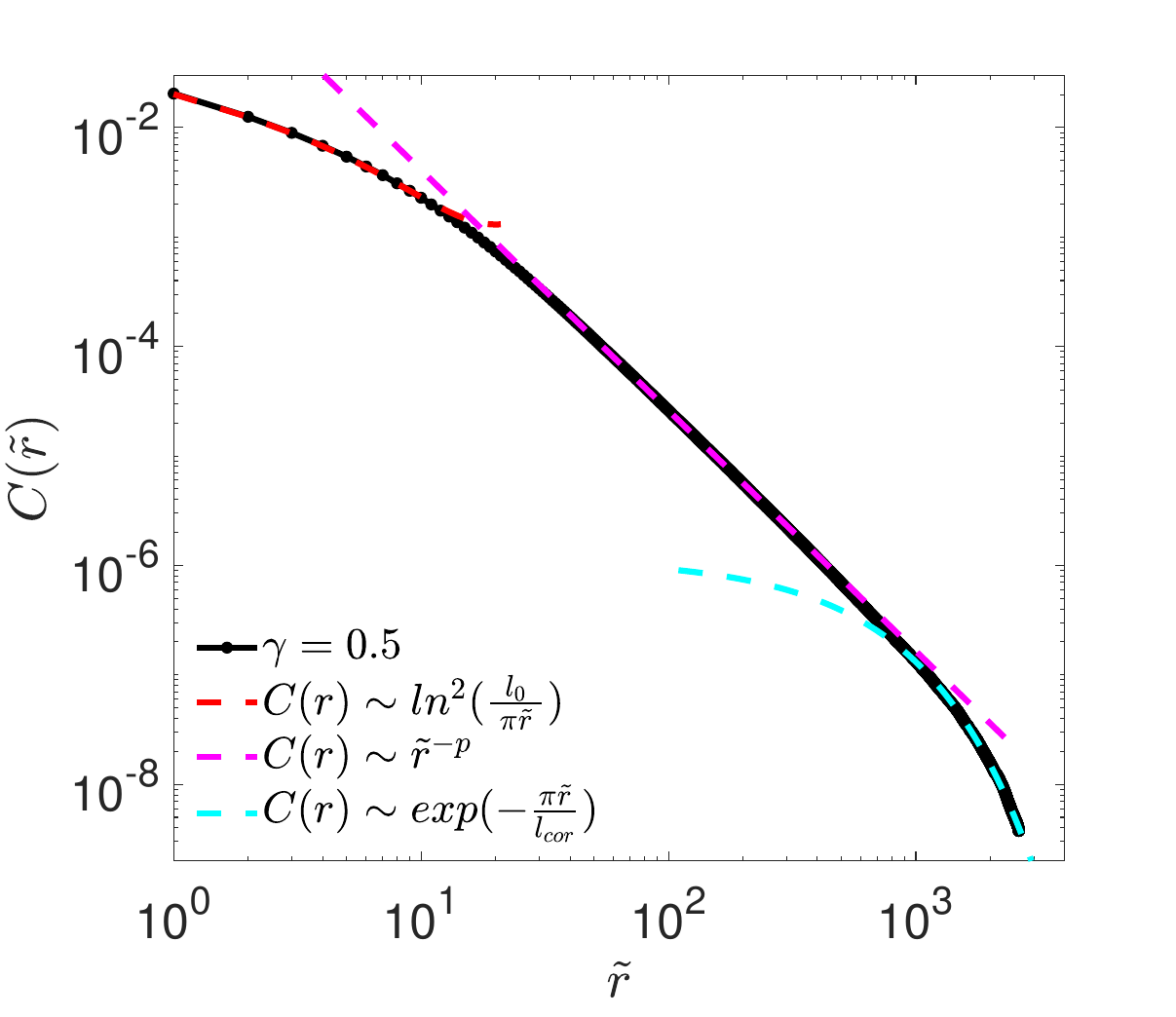} \label{fig:PM_Cr_fit}}\\
	\subfigure{\includegraphics[width=7cm]{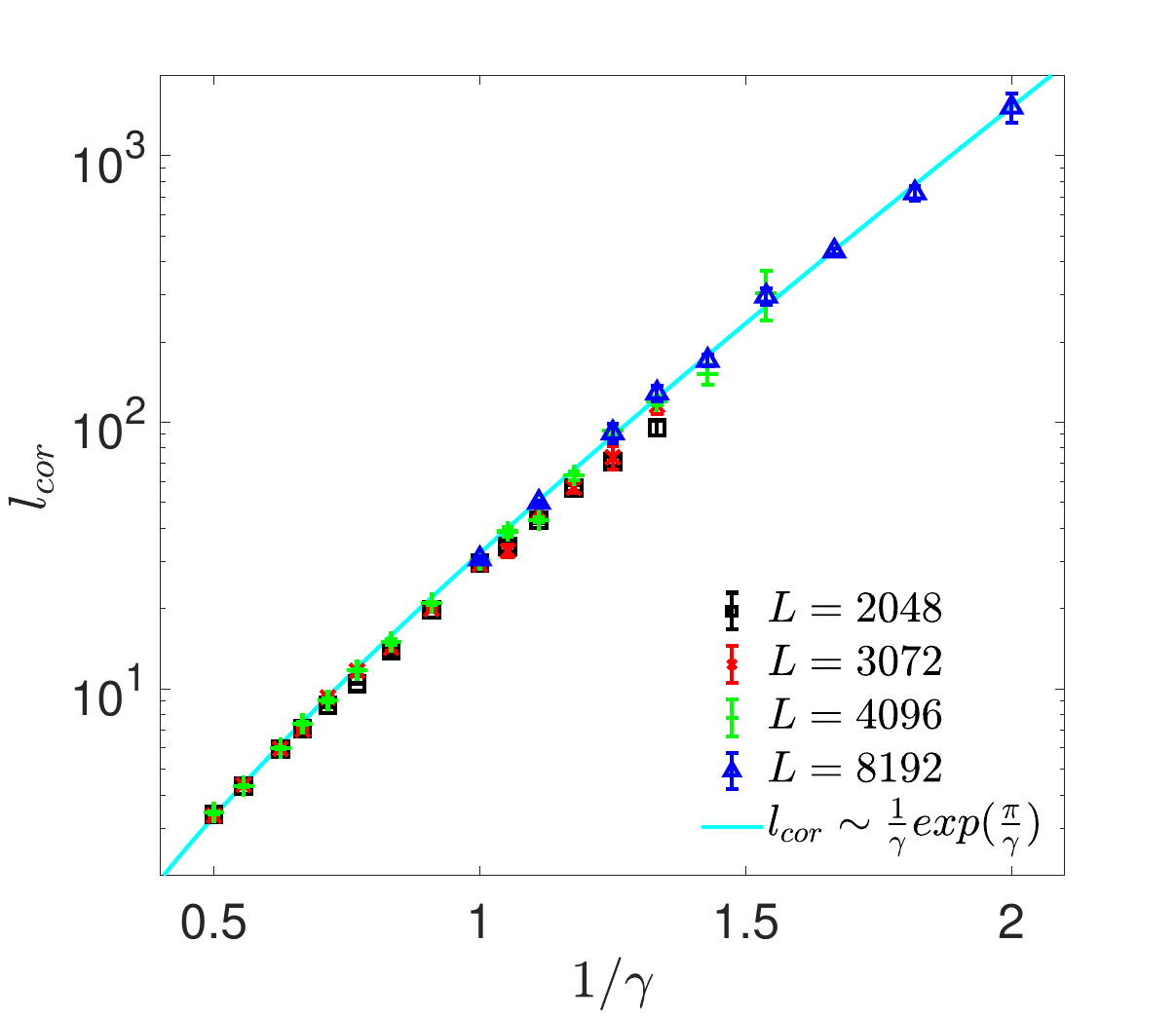} \label{fig:PM_lcor_fit}}
	\caption{Upper: $C(\tilde{r})$ in Eq.~(\ref{eq:Cr}) using the PM protocol, see Appendix \ref{app:protocol} for details, with $L = 8192$, $34$ trajectories and an additional time average over four equidistant points for $t \ge L/2$. 
	The dashed curves stand for the fittings to: ballistic ($\log^2$ decay), diffusive ($1/r^2$ decay) and exponential decay. The monitoring strength is $\gamma = 0.5$ and the fitting parameters are $l_0 \approx 63\pm 3$, $p \approx 2.20\pm0.0025$, $l_{\rm cor} \approx 1450 \pm 2$. Lower: The correlation length $l_{\rm corr}$ from the fitting of $C(r)$ as a function of $\gamma$. Depending on $\gamma$  and $L$ we average over both $30-100$ trajectories and $4-8$ equally spaced time points $t \in [L/2,L]$.} 
	\label{fig:PM_Cr_lcor_fit}
\end{figure}
Results depicted in Fig.~\ref{fig:PM_Cr_lcor_fit} (upper) show that the numerical $C(\tilde{r})$, where for convenience we drop the $t \to \infty$ dependence, is consistent with the analytical predictions of Ref.~\cite{poboiko2023}. 
For intermediate distances $l_0\ll r\ll l_{\rm cor} $, with $l_0\sim 1/\gamma$ the mean free path, $C(\tilde{r})$ follows a power-law decay $C(\tilde{r}) \sim \tilde{r}^{-p} $ with $p\sim 2$. However, for $r\gg l_{\rm cor}$, with $l_{\rm cor} > 1000$, the correlation function decays exponentially as $C(\tilde{r}) \sim \exp(-{\pi \tilde{r} \over l_{\rm cor}}) $. We stress that previous studies $L \lesssim 1000$ would have missed this exponential decay and therefore the existence of the area-law.  
 We note that we rescale the distance $\tilde{r} = {L\over \pi} \sin{\pi r\over L}$ due to the implementation of periodic boundary conditions. 

Since our focus is on the existence of a MIPT, we are interested in characterizing the regime in which $C(r)$ decays exponentially. For that purpose, we compute $l_{\rm cor}$ as a function of $\gamma$ by comparing $C(r)$ with the analytic expression Eq.~\eqref{eq:RG}. We note that due to the nature of the approximation leading to Eq.~\eqref{eq:RG}, the prefactors are not expected to be quantitatively correct so for our comparison they are just fitting parameters. 
Numerical results depicted in Fig.~\ref{fig:PM_Cr_lcor_fit} (lower) show excellent agreement with the analytical expression, particularly the exponential growth of $l_{\rm corr}$ with $1/\gamma$, which indicates the absence of a MIPT. 
Further evidence supporting this conclusion is provided in Appendix \ref{app:implargeL}, where we show explicitly the importance of reaching $L \sim 10000$ to rule out the existence of a MIPT. For smaller systems $L\le 4092$, finite-size effects can artificially produce a finite critical monitoring strength $\gamma_c \sim 0.1 $ using a fitting function $\propto \exp(a/|\gamma -\gamma_c|)$ with $a$ a constant, and $\gamma_c \sim 0.2$ for  Eq.~(\ref{eq:BKT}) mimicking the existence of a MIPT characterized by vortex unbinding. However, for $L = 8192$, $\gamma_c = 0.0 \pm 0.1$ using Eq.~(\ref{eq:RG}) and $\gamma_c = 0.06\pm 0.1$ using  Eq.~(\ref{eq:BKT}),  so $\gamma_c$ is consistent with zero, thus confirming the absence of a MIPT. \\
\subsection{Quantum State Diffusion Protocol} 
We now compute the EE using the QSD protocol, see Appendix \ref{app:protocol} for further details. We recall that in Ref.~\cite{cao2019a} it was predicted no MIPT based on a phenomenological model of the effect of measurements on the dynamics which was supported by numerical results for $L \leq 512$. By contrast, using 
semi-analytical arguments, and numerical results for $L \le 500$, a transition with 
\begin{equation}
	l_{\rm cor} \sim \exp \left({b\over\sqrt{|\gamma -\gamma_c|}}\right)
	\label{eq:BKT}
\end{equation}
and $\gamma_c \approx 0.31$ was predicted \cite{alberton2021a} . Because both protocols share the same symmetry, it has been predicted \cite{fava2024} the absence of MIPT in this case.  We show next that a numerical calculation of $C(r)$ for $L \leq 16384$ confirms this prediction. The computation time for this protocol is substantially shorter which enabled us to reach larger sizes in even less time than for the PM protocol, about one month and a half with two A100 GPU cards. 
 \begin{figure}
	\centering
  \includegraphics[width=7cm]{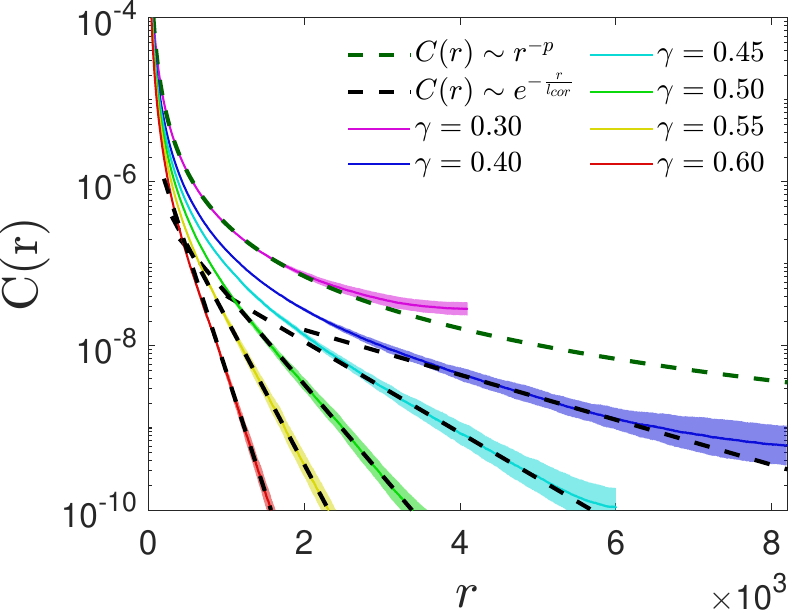}\\
  \includegraphics[width=6.8cm]{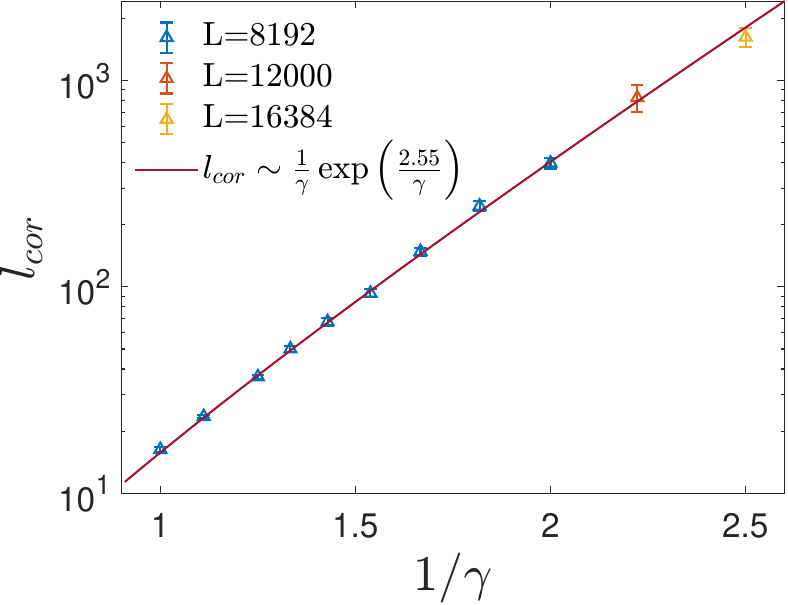}
  \caption{Upper: $C(r)$ in Eq.~(\ref{eq:Cr}) for the QSD protocol, different monitoring rates $\gamma$, $L = 16384$ for $\gamma=0.4$, $L=12000$ for $\gamma=0.45$, and $L=8192$ otherwise. The time step is $dt=0.05$. The width of each curve is the error bar. For each $\gamma$, we average over both at least $10$ trajectories and $33$ equally spaced time points $t\in [L/2,L]$ after saturation. For $\gamma\leq 0.5$, $C(r)\sim r^{-p}$ with, $p \sim 2$ ($p  = 2.1$ for $\gamma = 0.3$). 
  We need to reach $L = 12000, 16384$, much larger than in previous studies, to observe an exponential decay $\sim \exp(-r/l_{\rm cor})$ for $\gamma = 0.45, 0.4$ respectively. Lower: Correlation length $l_{\rm cor}$ from the fitting of $C(r)$ for the QSD protocol for different sizes $L$ compared with the analytic prediction Eq.~(\ref{eq:RG}). For instance, $l_{cor}=1620 \pm 165$ for $\gamma = 0.4$.
  	 Shifting $\gamma\rightarrow \gamma-\gamma_c$ and treating $\gamma_c$ as an additional fitting parameter yields $\gamma_c = 0.00 \pm 0.10$, which confirms the absence of a MIPT.} 
  \label{fig:Cr}
\end{figure}
 In Fig.~\ref{fig:Cr} (upper), we depict $C(r)$ for different monitoring rates $\gamma$. We observe that for $r$ sufficiently small, $C(r)\sim 1/r^2$, which corresponds to the {\it volume}-law phase with logarithmic growth of the EE. For $\gamma\geq 0.5$, and sufficiently large $L$, $C(r)\sim \exp(-r/l_{\rm cor})$ that characterizes \cite{poboiko2023} the area-law phase. For smaller $\gamma = 0.4,0.45$, in order to observe the exponential decay in $C(r)$, and compute $l_{\rm cor}$, it is necessary to reach $L = 12000$ and $L = 16384$ respectively.
 Therefore, the mentioned discrepancies in the literature were caused by the small lattice sizes $L \lesssim 1000$ considered.    
  We have checked that for $r < 2000$, the power-law decay region, and at $\gamma=0.4$, $C(r)$ for $L=8192$ is indistinguishable from that using those larger sizes, so it is unbiased to present in the same figure $l_{\rm cor}$ for different size $L$.

Next, we extract $l_{\rm cor}$ from the exponential decay of $C(r)$ for sufficiently large $r$, see Fig.~\ref{fig:Cr} (upper). 
More specifically, we obtain $l_{\rm cor}$ for each trajectory by a fitting of $C(r)$ to an exponential. We find that the resulting distribution is approximately Gaussian. The correlation length $l_{cor}$ equals to the mean of this distribution. The error bar of $l_{cor}$ is defined as the $95\%$ confidence interval of the distribution. 
 
The resulting error bars $\gamma = 0.4$ ($L=16384$) and $\gamma = 0.45$ ($L=12000$) with $\sim 10$ trajectories are larger because, at fixed $L$, and near the crossover from power-law to exponential decay, the exponential regime is narrow and close to the boundary. Moreover, trajectory-to-trajectory fluctuations are larger for smaller $\gamma$.

We then proceed to fit the obtained $l_{\rm cor}$ with Eqs.~(\ref{eq:RG}),  (\ref{eq:BKT}). We perform the fitting with Eq.~(\ref{eq:RG}) by shifting the exponent $\gamma\to \gamma -\gamma_c$ and considering $\gamma_c$ as a fitting parameter. In order to determine the error bar of $l_{\rm cor}$ more accurately, we use the weighted least squares method, that minimizes the residues, weighted by the inverse of the variance. 
The results of the fit, see Fig.\ref{fig:Cr} (lower) above and Fig.\ref{fig:QSD_gammaCfit} in Appendix \ref{app:implargeL}, show that both expressions describe the data well, yielding $\gamma_c = 0.00 \pm 0.10$  for Eq.~(\ref{eq:RG}), $\gamma_c = 0.04 \pm 0.08$ for Eq.~(\ref{eq:BKT}). Therefore, we conclude that there is no MIPT in 1d for the QSD protocol.\\
\section{Entanglement dynamics in 2d}
We now turn to the study of MIPT in 2d by using exactly the same Hamiltonian and protocols but defined on a square lattice. 
Previous results employing the PM \cite{poboiko2023a}, see also \cite{fava2024}, and QSD \cite{chahine2024} protocols found a MIPT  at a finite $\gamma$. However, the reported MIPT for each protocol is qualitatively different. According to Ref.~\cite{chahine2024}, the transition in the QSD protocol occurs in the {\it volume}-law phase and the EE is the natural scaling observable. By contrast, in Ref.~ \cite{poboiko2023a} it was found analytically that 
 the mutual information remains scale invariant at the MIPT while numerical results indicate that the transition occurs in the area law phase. 
 
 \begin{figure}[!htbp]
	\centering
	\subfigure{\includegraphics[width=6.7cm]{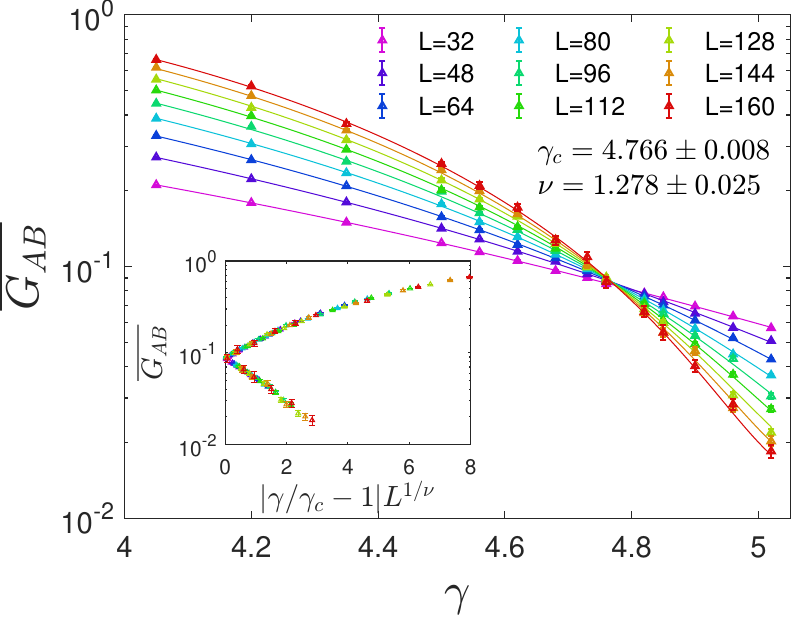}}\\ %\label{fig:CrossQSD}}
	\subfigure{\includegraphics[width=6.7cm]{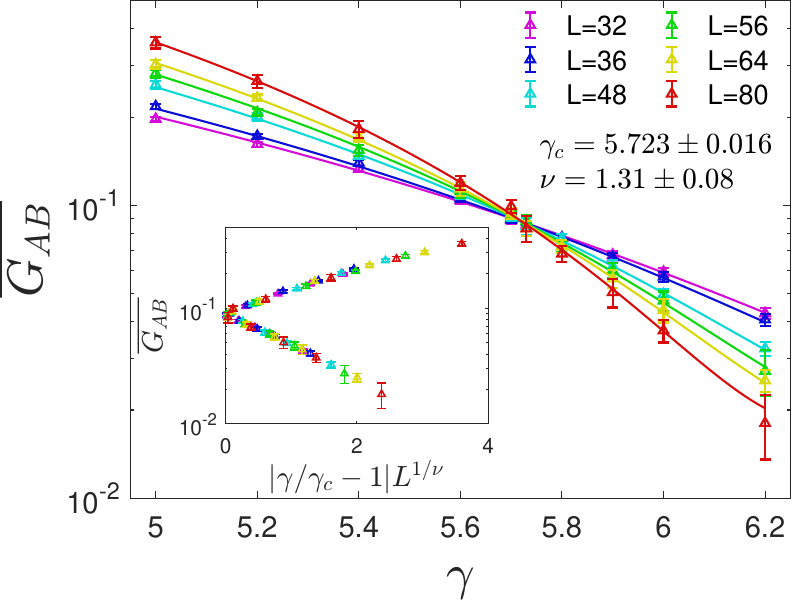}} %\label{fig:CrossPM}}
	\caption{Particle number covariance $\overline{G_{AB}}$ in Eq.~(\ref{eq:GAB}) as a function of $\gamma$ for different sizes $L$. The lines stand for least‐squares polynomial fittings, see \cite{Ohtsuki2014} and main text for details. Upper: QSD protocol. A sharp crossing occurs at $\gamma_c \approx 4.77 \pm 0.01$ indicating the existence of a MIPT characterized by a scale invariant $\overline{G_{AB}}$. 
The inset depicts the optimal data collapse achieved by rescaling  $\overline{G_{AB}}$ with  $\lvert \gamma/\gamma_c - 1\rvert\,L^{1/\nu}$ around $\gamma_c$, where $\nu \approx 1.28 \pm 0.03$ is the critical exponent governing the divergence of the correlation length. 
Lower: PM protocol. Similarly, the crossing is at $\gamma_c \approx 5.72 \pm 0.02$, (note that the most of the disagreement in $\gamma_c$ with respect to Ref.~\cite{poboiko2023a} is due to a different definition of the waiting times needed to compare both protocols) the optimal data collapse (inset) is at $\gamma_c = 5.72 \pm 0.02$ and $\nu = 1.31 \pm 0.08$.  The value of $\nu$ agrees with the analogue for the 3d non‑Hermitian Anderson transition \cite{Shindou2021}. Note that \cite{poboiko2023a} $\mathcal{I}_2 \simeq {2\pi^2\over  3}\overline{G_{AB}}$ for $\gamma \lesssim \gamma_c$.  }
	\label{fig:2Dcross}
\end{figure}
 In order to identify and characterize the MIPT, we study the mutual information $\mathcal{I}_2 = S_A + S_B -S_{A \cup B}$  in the steady state as a function of the monitoring strength $\gamma$ for different lattice sizes $L$. For that purpose, we use  \cite{poboiko2023a,levitov2009} $\mathcal{I}_2 \simeq {2\pi^2\over 3}\overline{G_{AB}}$ where $\overline {G_{AB}}$ is the particle covariance $\overline {G_{AB}}$ between two regions $A$ and $B$, computed \cite{levitov2009} from $C(\vec{r})$,
 \begin{equation}
   \overline{G_{AB}}=-\int_A d^2 \vec{r_1} \int_B d^2 \vec{r_2} C(\vec{r_1}-\vec{r_2}) \quad \vec{r_1}\in A, \vec{r_2}\in B,
   \label{eq:GAB}
 \end{equation}
 \begin{figure}
 	\centering
 	\subfigure{\includegraphics[width=6.4cm]{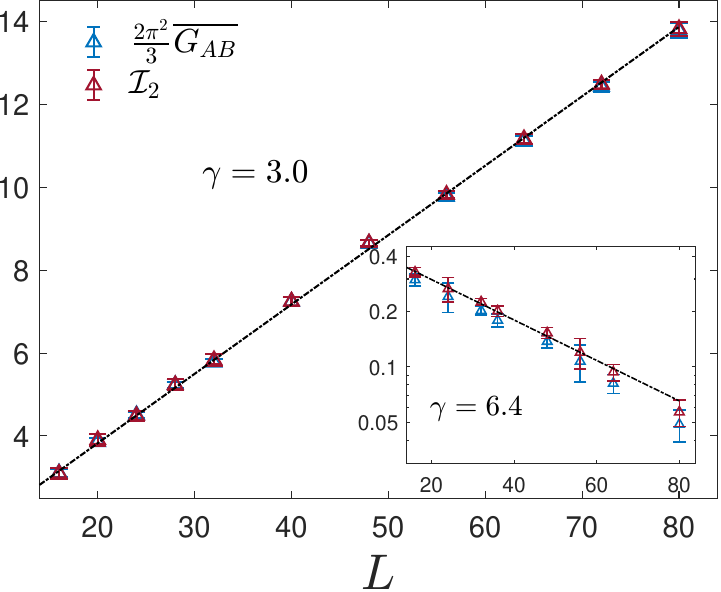}} %\label{fig:I2volarera}}
 	\subfigure{\includegraphics[width=6.6cm]{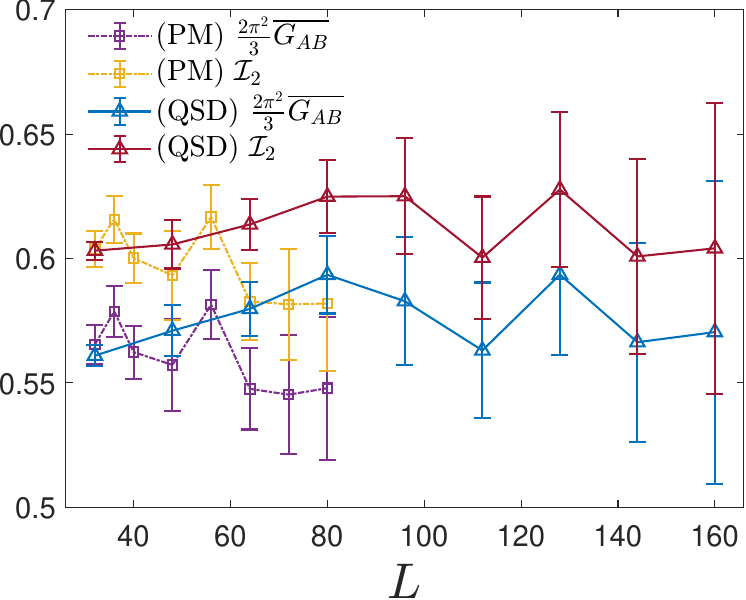}} %\label{fig:I2MIPT}}
 	\caption{Mutual information $\mathcal{I}_2$ and the particle‐number covariance $\overline{G_{AB}}$ in Eq.~(\ref{eq:GAB}) versus system size $L$. 
 		Upper: PM protocol: $\gamma  = 3$, $\mathcal{I}_2 \simeq {2\pi^2\over 3}\overline{G_{AB}} \propto L$ so the system is in the {\it volume}-law phase. Inset: $\gamma = 6.4$, both $\mathcal{I}_2$ and $\overline{G_{AB}}$ decays exponentially so the system is in the area-law phase. Lower:  $\mathcal{I}_2$ and $\overline{G_{AB}}$ at $\gamma_c$ do not depend on $L$ for both PM and QSD protocols. Moreover, its value at the MIPT does not depend much on the protocol. }
 \label{Fig:Iac_GAB_vs_L}
 \end{figure}
where we choose $A$, $B$ to be $L\times L/4$ blocks, separated by another $L\times L/4$ block. We note that though this relation is perturbative \cite{levitov2009}, we shall show later, see Fig.~\ref{Fig:Iac_GAB_vs_L} (upper), that it is still quantitatively valid for $\gamma \lesssim \gamma_c$. For larger $\gamma$, the prefactor gradually increases. For smaller sizes using the PM protocol, this was confirmed earlier in Ref.~\cite{poboiko2023a}.
  As in the 1d case, we perform averages over both different trajectories and equally spaced time points in the interval $t\in [L/2,L]$. Comparatively, the EE saturates much faster 
  than in the 1d case, enabling us to reach size $160\times160$ for the QSD protocol. VRAM usage is the limitation to reach even larger sizes. 
  For the PM protocol, since the monitoring rate is large, we need an exponentially larger number of measurements per unit of time, so we can only reach $80\times80$, which is still much larger than in previous studies ($\leq 40\times40$)  \cite{poboiko2023a}.
   Results for $\overline{G_{AB}}$, see Fig.~\ref{fig:2Dcross}, as a function of $\gamma$ for different $L$ show a sharp crossing at a certain protocol dependent $\gamma = \gamma_c$, which indicates the existence of a MIPT characterized by a $L$ independent mutual information \cite{poboiko2023a}. Therefore, for $\gamma \approx \gamma_c$, entanglement is different from that in the area-law and {\it volume}-law phases. 

To characterize the critical phase near $\gamma_c$, we perform a finite‐size scaling analysis by defining the scaling variable $\phi_1 = u(w)L^{1/\nu}$, where $w = (\gamma - \gamma_c)/\gamma_c$ and $u(w)=\sum_{i=1}^m b_i w^i$. The mutual information $\mathcal{I}_2$, and hence $\overline{G_{AB}}$, can then be expressed as a polynomial in $\phi_1$, namely $\overline{G_{AB}}=\sum_{i=0}^n a_i \phi_1^i$ \cite{Ohtsuki2014}.
In practice, we fit with polynomial orders $m=2$ and $n=3$, fixing $a_1=1$ and treating the crossing point $\gamma_c$ and exponent $\nu$ as fit parameters (seven parameters in total), using least‐squares regression to minimize the residue.
 The maximum polynomial order in $w$ is $m\times n = 6$, which is necessary to account for the power‐law decay when $\gamma < \gamma_c$ and the exponential decrease when $\gamma > \gamma_c$. 
 
 For the QSD protocol, we have more data points so we perform a more refined analysis by introducing a second field, $\phi_2 \sim L^{-\alpha}$ ($\alpha > 0$) that captures the contribution of irrelevant scaling. The corresponding expansion of the scaling function $\overline{G_{AB}}$ modifies as $ \overline{G_{AB}}(\phi_1,\phi_2)=\sum_{i_1=0}^{n_1}\sum_{i_2=0}^{n_2} a_{i_1,i_2} \phi_1^{i_1}\phi_2^{i_2} $,
   where we fix $a_{0,1}=a_{1,0}=1$, and the choice $m=2$, $n_1=3$, and $n_2=1$ yields a total of $12$ fitting parameters. The fitted irrelevant exponent is $\alpha \approx 3$, so the corresponding correction decays rapidly with increasing $L$. This explains the very small crossing shift at small sizes $L=32,48$ in Fig.~\ref{fig:2Dcross} (upper) and indicates that the anomaly vanishes for sufficiently large $L$. 
   
   The fitting results show that $\gamma_c$ is protocol dependent: $\gamma_c \approx 4.77 \pm 0.01$ for the QSD and $\gamma_c = 5.72\pm 0.02$ for the PM protocol. The critical exponent $\nu$, which controls the divergence of the $l_{\rm cor}$ as we approach the transition from the area-law, is similar for both protocols: for QSD, $\nu = 1.28 \pm 0.03$; for PM, $\nu = 1.31 \pm 0.08$. In the insets of Fig.~\ref{fig:2Dcross}, we show that, rescaling the data using $|\gamma/\gamma_c - 1|\,L^{1/\nu}$, results for different system sizes $L$ collapse onto a single line for both $\gamma<\gamma_c$ and $\gamma>\gamma_c$.   
   We note that $\nu\sim 1.3$ is different from both the analytic NLSM result $\nu = 1$ and the numerical result for the Anderson transition in the BDI class $\nu = 1.06\pm 0.02$ \cite{wang2021}. We cannot rule out that the symmetry controlling $\nu$ is that of effective non-Hermitian Hamiltonian \cite{xiao2025} which belongs to class AI$^\dag$. We also find agreement with the critical exponent $\nu \in [1.276,1.288]$ in the 3D non‑Hermitian Anderson model \cite{Shindou2021} belonging to this AI$^\dag$ class.  
   
 Finally, we study the $L$ dependence of $\mathcal{I}_2$ and check that $ \mathcal{I}_2 \approx \frac{2\pi^2}{3} \overline{G_{AB}}$ \cite{poboiko2023a}. Results depicted in Fig.~\ref{Fig:Iac_GAB_vs_L} show that $\mathcal{I}_2$ is $L$ independent at the MIPT, linear in $L$ in the {\it volume}-law phase, and decays exponentially with $L$ in the area-law phase. $\mathcal{I}_2$ and $\overline{G_{AB}}$ differ at most by an overall constant so either observable can be used to characterize the transition. Moreover, there are no substantial differences between the two protocols.  \\
\section{Conclusions}
Exploiting the advantages of GPU computing, we have shown that the EE of 1d Dirac fermions with the occupation number being monitored using PM and QSD protocols is always in the area-law phase.
In 2d, we have identified a MIPT at a protocol-dependent critical monitoring strength characterized by a size independent mutual information so the entanglement properties around the transition are different from those in the area-law and {\it volume}-law phases. The critical exponent that characterizes the approach to the MIPT is similar for both protocols $\nu \approx 1.3$ and agrees with that of the 3d non-Hermitian Anderson transition in class AI$^\dagger$.
The absence of the MIPT in 1d but its existence in 2d, together with the scale invariance of the mutual information at the MIPT were predicted earlier by the NLSM \cite{poboiko2023a,fava2024}. However, the NLSM does not predict correctly the value of the critical exponent or the critical monitoring. Therefore, for a quantitative description of the entanglement dynamics a numerical approach capable of simulating large lattices is required. 
\vspace{0.5cm}
\\
\vspace{0.5cm}
\centerline{\bf Acknowledgments}
A. M. G. thanks Yan Fyodorov, Kohei Kawabata and Marco Schiro for illuminating discussions. We thank Zhenyu Xiao for interesting conversations. We acknowledge support from the National Natural Science Foundation of China (NSFC): Research Fund for International Senior Scientists No. 12350710180, Individual Grant No. 12374138. B.F. acknowledges support from the China Postdoctoral Science Foundation (Grant numbers: 2023M732256, 2023T160409, GZB20230420).
\setcounter{secnumdepth}{1}   
\appendix
\section{PM and QSD protocols}\label{app:protocol}
This appendix introduces the numerical details of the measurement protocols. 
For the PM protocol, we initialize the system in the N\'{e}el state $|\psi(t=0)\rangle = |10101010\ldots\rangle$ and evolve the monitored system using the corresponding correlation matrix $D_{ij}(t) = \langle \psi(t)| c_i^\dagger c_j |\psi(t)\rangle $. For numerical stability and efficiency, the time evolution is performed directly on this correlation matrix. Between two successive measurements, the system evolves unitarily under the Hamiltonian $H$. The updated correlation matrix is given by $D (t+\tau) = e^{-iH\tau} D(t) e^{iH\tau} $, where $\tau = {\rm ln}(\eta)/(\gamma N)$ and $\eta \in (0,1]$ is a uniformly distributed random number. $N=L/2$ is the particle number and $\gamma$ is the measuring rate per site.
At each measurement step, namely, at time $t+\tau$, we first randomly select a site $j$, and compute the local occupation number $p_j = D_{j,j}$. We then generate a random value $p_c \in[0,1]$. If $p_j\ge p_c$, we perform the projection using the operators $ \hat{P}_1(j) = \hat{n}_j $. The corresponding correlation matrix is updated following
\begin{equation}
	D_{i,i'} = \langle \psi' | c_i^\dagger c_{i'} | \psi' \rangle
	= \begin{cases}
		1, ~i=i'=j \\
		0,  (i=j, i\ne i') ~\text{or}~ (i'=j, i\ne i') \\
		\langle c_i^\dagger c_{i'} \rangle - \frac{\langle c_j^\dagger c_{i'} \rangle \langle c_i^\dagger c_j \rangle}{\langle c_j^\dagger c_j \rangle}, ~ \text{\rm otherwise}.
	\end{cases}
	\label{eq:QJ_update}
\end{equation}

If $p_j< p_c$, we perform the projection operation $ \hat{P}_0(j) = 1 - \hat{n}_j $. The corresponding updated correlation matrix elements are given by:
\begin{equation}
	D_{i,i'} = \langle \psi' | c_i^\dagger c_{i'} | \psi' \rangle
	= \begin{cases}
		0, ~~~~~ (i=j) ~\text{or}~ (i'=j) \\
		\langle c_i^\dagger c_{i'} \rangle - \frac{\langle c_j^\dagger c_{i'} \rangle \langle c_i^\dagger c_j \rangle}{1 - \langle c_j^\dagger c_j \rangle}, ~ \text{otherwise}
	\end{cases}
	\label{eq:PM_update_0}
\end{equation}

After each measurement, we reconstruct the new orthonormal state $|\psi(t=0)\rangle$ from the correlation matrix $D$ using the singular value decomposition $ D = U S U^\dagger $, where $ S_{i,i} = 1~(1 \le i \le N ) $ and $ 0~(N + 1 \le i \le L) $. The new state matrix $ U $ characterizes the new state $ |\psi(t+\tau) \rangle $ at time $ t + \tau $.
We repeat the calculation until $t \geq L/2$ to ensure that the system reaches the steady state at which the EE does not experiences any net growth.

In the QSD protocol, we likewise initialize the system in the N\'{e}el state $|10101010\ldots\rangle$ and evolve coefficient matrix $U$ of the wave function in the particle-number basis, which is defined as:
\begin{equation}
	|\psi(t)\rangle = \prod_{k=1}^{N} \left[\sum_{j=1}^L U_{jk}(t) c_j^\dagger\right] | \text{vac} \rangle
\end{equation}
where $|\text{vac} \rangle$ is the vacuum state annihilated by any $c_i, i=1,2\cdots, L$; $U$ is an $L\times N$ matrix satisfying $U^\dag U=\mathbf{1}_{N\times N}$, where $\mathbf{1}_{N\times N}$ denotes the $N\times N$ identity matrix.
Unlike the PM protocol, in which measurement events are performed randomly in space and time, here we continuously measure the system uniformly in space and time by introducing independent Gaussian random noise $dW^t_i$ with zero mean at each site $i$, satisfying $d W^t_i d W^t_j =\delta_{ij} \gamma dt$, $\gamma$ is the measurement rate. The measurement term $i\sum_i dW^t_i \hat{n}_i$, is non-Hermitian, so that the system evolves non-unitarily by the state evolution equation:
\begin{equation}
\begin{aligned}
U(t+dt) \propto \exp\biggl(&-i\tilde{H} dt+d W^t \\
&+(2\langle \hat{n}\rangle_t-\mathbf{1}_{L\times L})\gamma dt\biggr) U(t)
\end{aligned}
	\label{eq:QSD_evo}
\end{equation}
where $\tilde{H}_{ij}=J\delta_{i\pm 1,j}$ is the $L\times L$ coefficient matrix of the Hamiltonian $H$, namely, $H=\sum_{ij}\tilde{H}_{ij}c_i^\dag c_j$. $d W^t$ and $\langle \hat{n}\rangle_t$ are the $L\times L$ diagonal matrix whose $i$-th diagonal element equals to $d W^t_i$ and $\langle \hat{n}_i\rangle_t=\sum_j U_{ij}U_{ij}^*$ respectively.
In practice, we use the fourth-order Runge–Kutta method to evolve the matrix $U(t)$ with a time step of $dt=0.05$, and we have verified that for $dt \le 0.1$, the results are identical to those obtained by exact diagonalization. After each step $t\rightarrow t+d t$,  we perform a QR decomposition on the matrix U, namely, we write U$=$QR with Q an unitary matrix and R an upper triangular matrix, to maintain the orth-normalization $U^\dag U=\mathbf{1}_{N\times N}$. As in the PM protocol, we repeat the calculation until $t \geq L/2$ when the system reaches a steady state. For $t \ge L/2$, we compute the density–correlation function and the entanglement entropy from the correlation matrix $D(t)=U(t)U(t)^\dag$. On a single A100, evolving one trajectory over $t\in [0,L]$ requires approximately 11 hours for $L=8192$ and $130$ hours for $L=16384$.
\section{Importance of large sizes $L \gtrsim 10000$ to demonstrate the absence of MIPT in 1d}\label{app:implargeL}
\begin{figure*}[t]
	\centering
	\includegraphics[width=6.5cm]{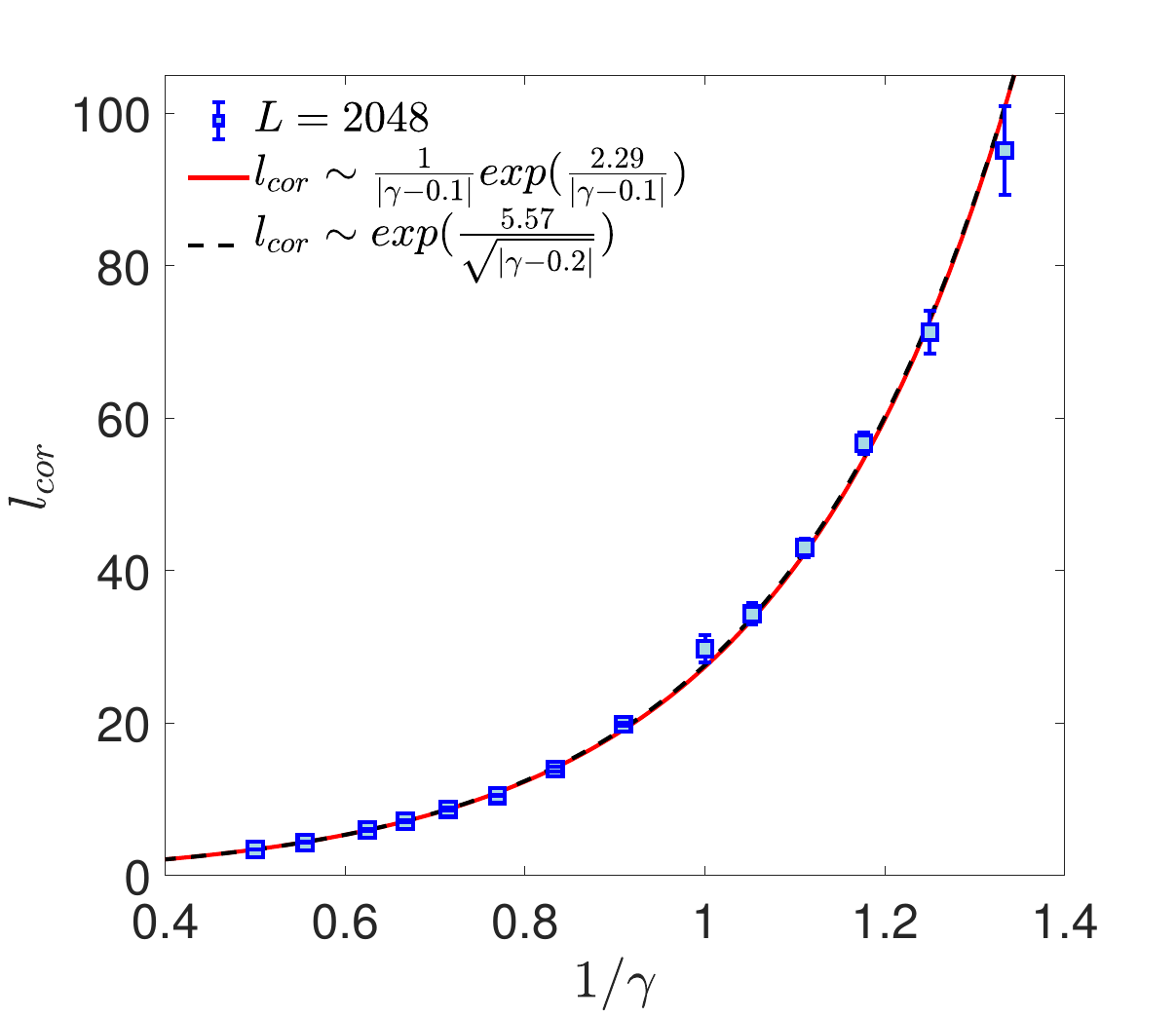}
	\includegraphics[width=6.5cm]{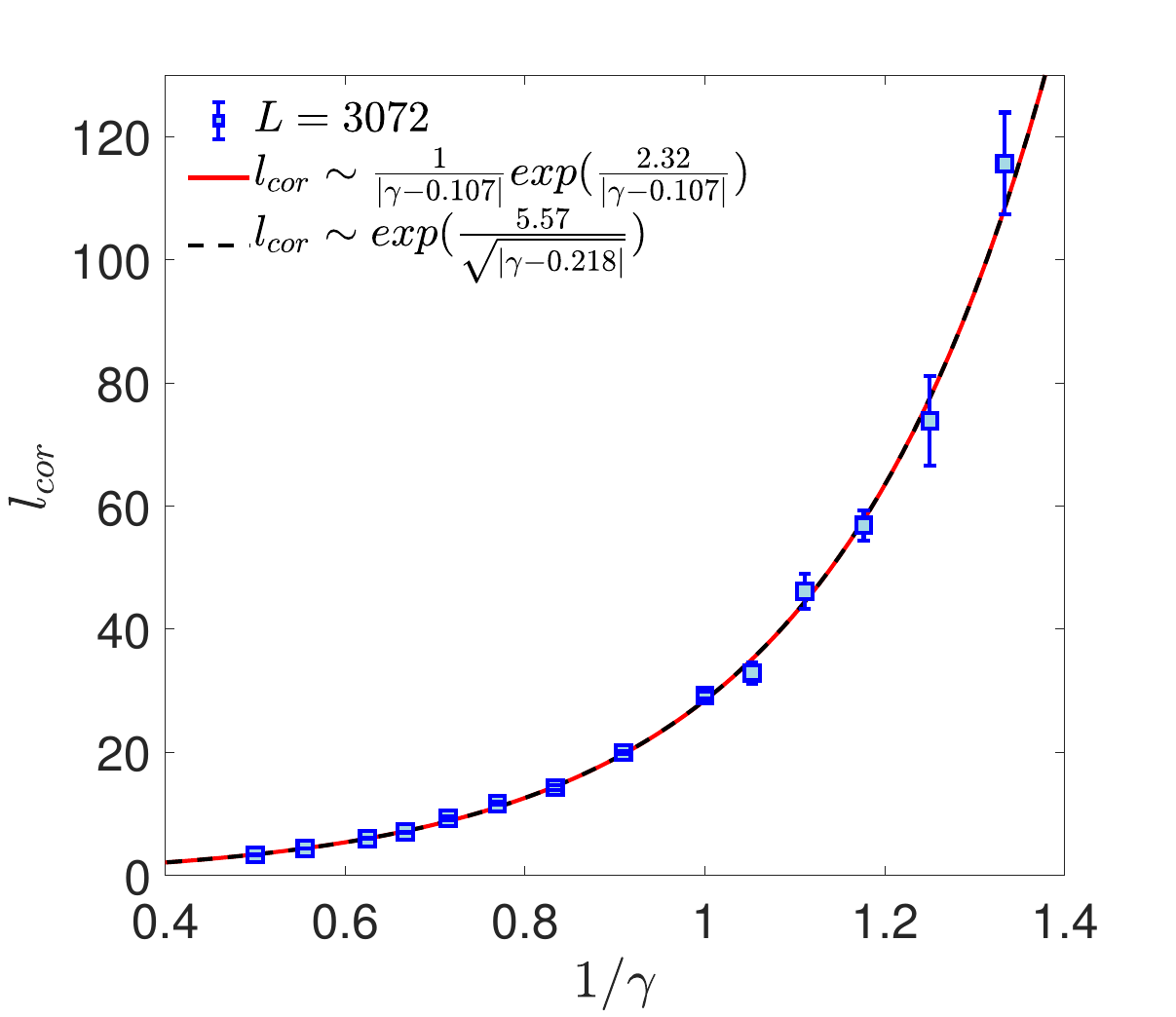}\\
	\includegraphics[width=6.5cm]{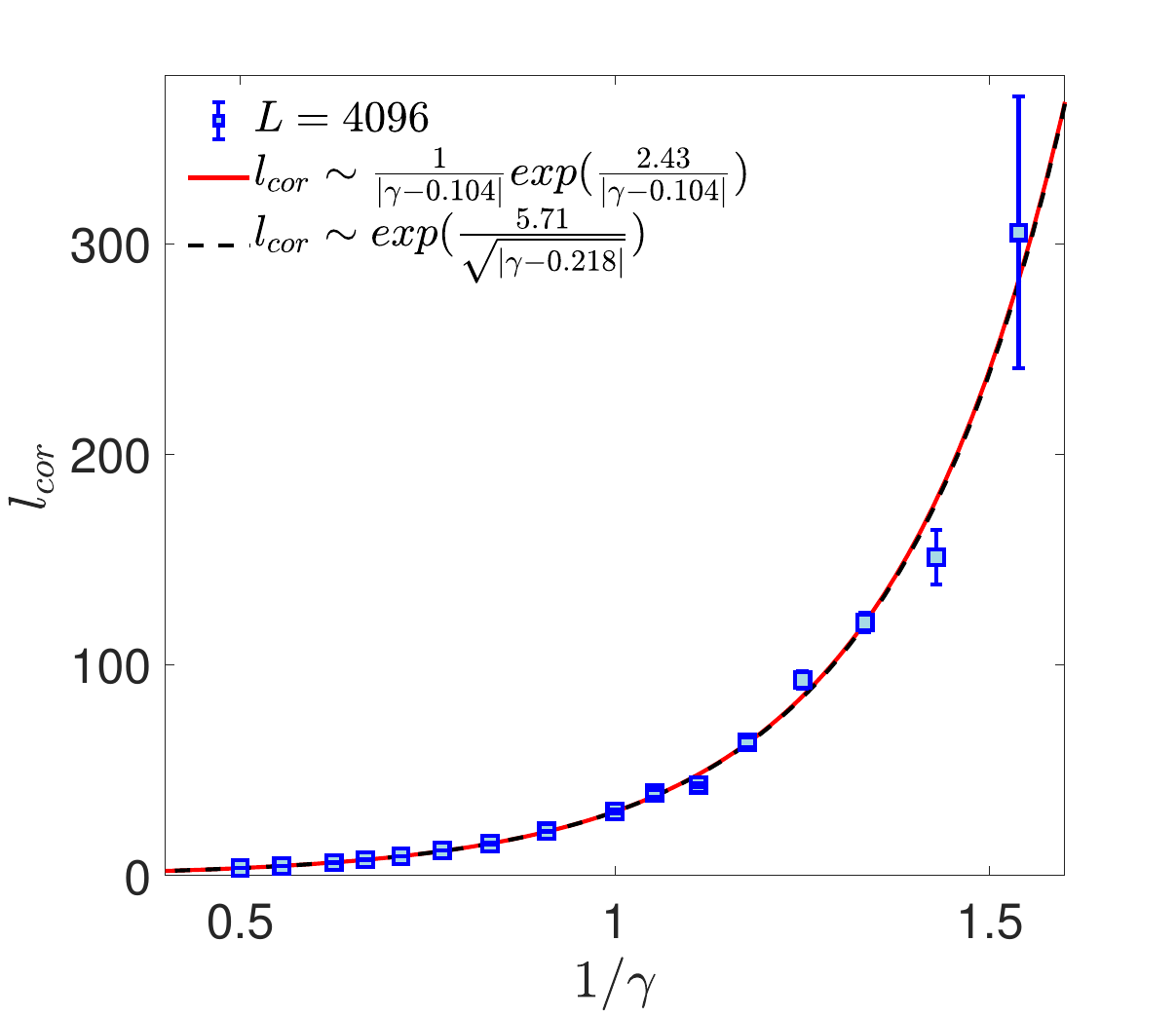}
	\includegraphics[width=6.5cm]{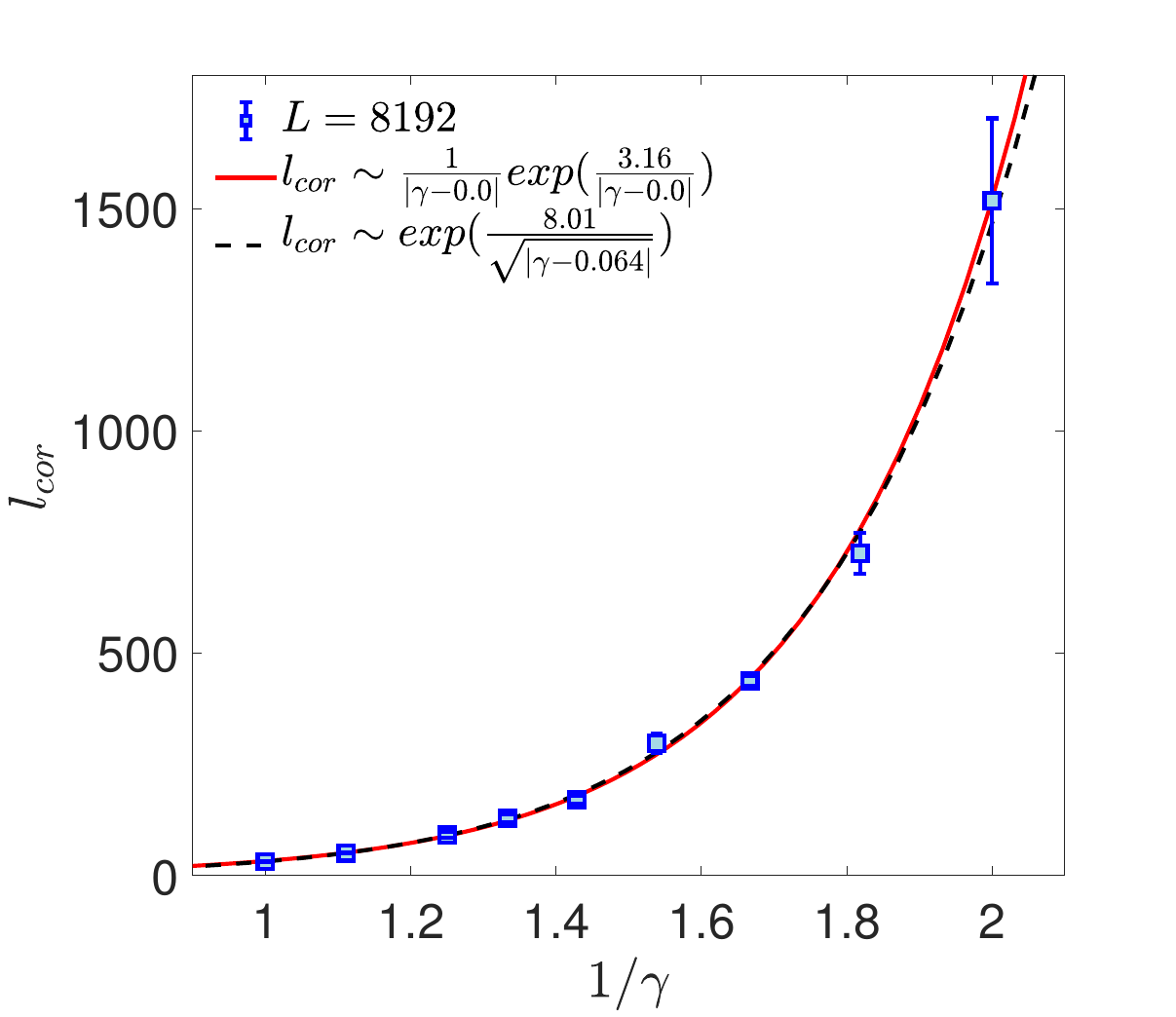}
	\caption{The correlation length $l_{\rm cor}$ resulting from the fitting of the exponential decay $C(r)$ as a function of $\gamma$. To extract $l_{\rm cor}$ from $C(r)$ when $\gamma \ge 0.6$, we only fit the region when $C(r)$ decays exponentially, but with $r \ll L/2$ to get rid of the boundary effects. For $\gamma \le 0.55$, we rescaled the distance following the details introduced in the main text. 
		We perform the fitting with both $l_{\rm cor} \sim {1\over |\gamma - \gamma_c|} \exp({a\over |\gamma - \gamma_c|})$ (Eq.~(\ref{eq:RG})) \cite{poboiko2023} and the BKT prediction (Eq.~(\ref{eq:BKT})) \cite{alberton2021a}  $l_{\rm cor} \sim \exp({b\over \sqrt{|\gamma - \gamma_c|}})$. For size $L = 4096$, we get a critical monitoring strength $\gamma_c = 0.1\pm 0.1$ for Eq.~(\ref{eq:RG}) and $\gamma_c = 0.22\pm 0.07$ for the BKT prediction. Similar results, suggesting a non-zero $\gamma_c$, occur for the BKT fitting function and smaller $N$. However, for size $L = 8192$, we obtain a critical monitoring strength $\gamma_c = 0.0 \pm 0.1$ using Eq.~(\ref{eq:RG}) and $\gamma_c = 0.06\pm 0.1$ for the BKT prediction Eq.~(\ref{eq:BKT}).}
	\label{fig:PM_lco_fit}
\end{figure*}

As a further verification of the absence of a MIPT in 1d, we analyze the behavior of the correlation length $l_{\rm cor}$ as a function of monitoring strength by varying the system sizes. This illustrates the importance of reaching large sizes in order to provide convincing evidence of the existence of a finite correlation length, and therefore no transition, for any finite monitoring strength. Specifically, we consider two fitting functions $l_{\rm cor} \sim {1\over |\gamma - \gamma_c|} \exp({a\over |\gamma - \gamma_c|})$ and the BKT prediction \cite{alberton2021a} $l_{\rm cor} \sim \exp({b\over \sqrt{|\gamma - \gamma_c|}})$, where $\gamma_c$
is a fitting parameter, restricted to $\gamma_c \ge 0$ and interpreted as the critical monitoring strength.
Our analysis focused on system sizes $L\ge 2048$, which allow us to reliably extract correlation lengths $l_{\rm cor}$ at the order of $100$ or larger under a relatively weak monitoring. Such large correlation lengths are essential for robust fitting and meaningful conclusions. 
For smaller system size, we need go to stronger monitoring strength to get a correlation length $l_{\rm cor}$. The resulting smaller correlation lengths make the data more easily compatible with a broad range of the fitting parameter, making it difficult to resolve the existence of a finite $\gamma_c$.
Moreover, our results confirm that in the strong monitoring regime, $l_{\rm cor}$ becomes independent of system size. Only in the weak monitoring limit, where we could obtain a larger $l_{\rm cor}$, we can meaningfully confirm whether a finite $\gamma_c$ exists.
A reliable extraction of $l_{\rm cor}$ in this regime requires substantially larger system sizes. For example, when system size $L= 2048$, we only get $l_{\rm cor} \sim 100$ for $\gamma \sim 0.7$, whereas for $L=8192$, we could get one magnitude larger $l_{\rm cor} \sim 1500$ under weaker monitoring $\gamma \sim 0.5$.

\begin{figure}[tbp]
	\centering
	\includegraphics[width=7cm]{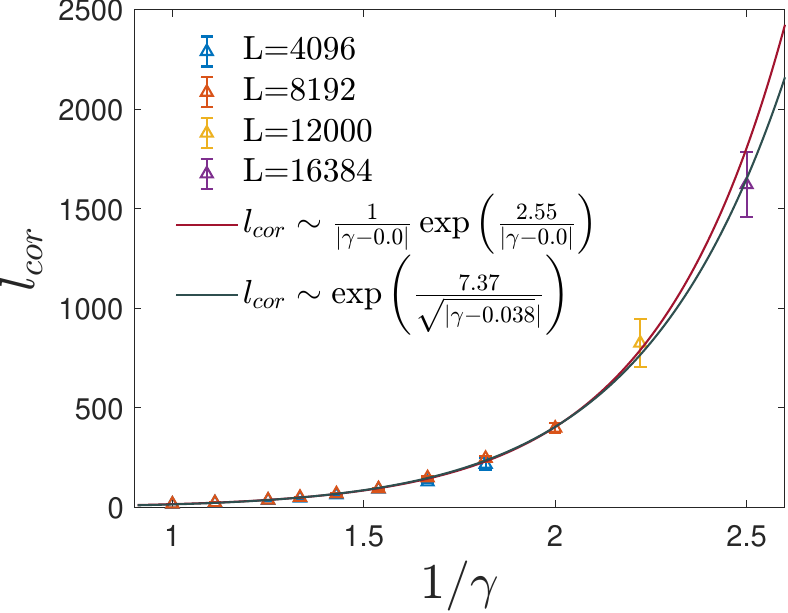}
	\caption{For QSD protocol, the correlation length $l_{\rm cor}$ resulting from the fitting of the exponential decay $C(r)$ as a function of $\gamma$, for both the analytical formula in Eq.~(\ref{eq:RG}) \cite{poboiko2023} and the BKT prediction in Eq.~(\ref{eq:BKT}) \cite{alberton2021a} with a shift $\gamma\rightarrow \gamma-\gamma_c, \gamma_c>0$. We fit the data in the range $\gamma \in [0.4, 0.8]$ of size $L=8192 (\gamma\in [0.5,0.8]),12000 (\gamma=0.45),16384 (\gamma=0.4)$, and we discard the strong $\gamma=0.9, 1.0$ that the formulas may not strictly apply. We also include the results for $L=4096$ to show the finite-size effects. We obtain a critical monitoring strength $\gamma_c = 0.00 \pm 0.10$ using Eq.~(\ref{eq:RG}) and $\gamma_c =  0.038 \pm 0.078$ for the BKT prediction in Eq.~(\ref{eq:BKT}).}
	\label{fig:QSD_gammaCfit}
\end{figure}

The results are shown in Fig.~\ref{fig:PM_lco_fit}.
For system size $L\le 4096$, the best fits suggest an apparent critical monitoring around $\gamma_c \sim 0.1$ and $\gamma_c \sim 0.2$ for the two employed fitting functions (see caption). However, as the system size increases to $L = 8192$, the fitted value of $\gamma_c$ approaches zero.
In Fig.~\ref{fig:QSD_gammaCfit}, we fit the data of the QSD protocol to the analytical formula Eq.~(\ref{eq:RG}) \cite{poboiko2023} and to the BKT prediction Eq.~(\ref{eq:BKT}) \cite{alberton2021a}, treating the critical value $\gamma_c>0$ as an additional (third) fitting parameter. As with the PM protocol, for the large size  we obtain a similarly small critical value: $\gamma_c = 0.0 \pm 0.1$ for Eq.~(\ref{eq:RG}) and $\gamma_c = 0.04\pm 0.08$ for the BKT prediction Eq.~(\ref{eq:BKT}). We also include the results for small system size $L=4096$, observing minor deviations from the $L=8192$ data, which can be attributed to finite-size effects. We find that for $L=16384$ at small $\gamma=0.4$, the data deviate from the exponential fit due to boundary effects: the fitting interval ($r\in[4250,6000]$) is narrow and the exponential-decay trend is blurred.

Considering all those analysis, we have provided compelling evidence against the existence of a MIPT, namely, for sufficiently large system size $L$, the system is in the area-law phase for any finite monitoring strength. 

\bibliography{monitoring.bib}

\end{document}